\documentclass[10pt,letter]{emulateapj}

\newcommand{\beq}{\begin{equation}}
\newcommand{\eeq}{\end{equation}}
\newcommand{\beqy}{\begin{eqnarray}}
\newcommand{\eeqy}{\end{eqnarray}}

\shorttitle{Quark deconfinement in Neutron stars}
\shortauthors{Staff et al.}

\begin{document}


\title{Quark deconfinement in neutron star cores:\\
    The effects of spin-down}


\author{Jan E. Staff \& Rachid Ouyed }
\affil{Department of Physics and Astronomy, University of Calgary,
2500 University Drive NW, Calgary, Alberta T2N 1N4, Canada}

\email{jstaff@phas.ucalgary.ca}

\and

\author{Prashanth Jaikumar} \affil{Physics Division, Argonne National
Laboratory, Argonne, Illinois 60439-4843, USA}

\begin{abstract}
We study the role of spin-down in driving quark
 deconfinement in the high density core of isolated neutron stars.
Assuming spin-down to be
solely due to magnetic braking, we obtain typical timescales to quark
deconfinement for neutron stars that are born with Keplerian
frequencies. Employing different equations of state (EOS), we
determine the minimum and maximum neutron star masses that will allow
for deconfinement via spin-down only. We find that the time to reach
deconfinement is strongly dependent on the magnetic field and that
this time is least for EOS that support the largest minimum mass at
zero spin, unless rotational effects on stellar structure are large.
For a fiducial critical density of $5\rho_0$ for the transition to the
quark phase ($\rho_0=2.5\times10^{14}$g/cm$^3$ is the saturation
density of nuclear matter), we find that neutron stars lighter than
$1.5M_{\odot}$ cannot reach a deconfined phase. Depending on the EOS,
neutron stars of more than $1.5M_{\odot}$ can enter a quark phase only
if they are spinning faster than about 3 milliseconds as observed now, whereas
larger spin periods imply that they are either already quark stars or
will never become one.

\end{abstract}


\section{Introduction}

Neutron stars, formed in the aftermath of core collapse supernovae,
are born rotating with spin periods that can be down in the
millisecond range. Coupled to this fast rotation, magnetic flux
conservation in the imploding progenitor provides the nascent neutron
star with magnetic fields in the range $10^{12}$-$10^{15}$G, as
confirmed by a variety of pulsar observations \citep{lyne}. A
customary first approximation is to model the pulsar as a rotating
magnetic dipole which loses angular momentum through electromagnetic
radiation. Consequently, the star spins down and the decreased
centrifugal force causes the core density to rise. As this occurs,
baryonic matter can undergo transitions to exotic phases of strongly
interacting matter, finally giving way to a phase in which quarks are
deconfined. Such a deconfined phase might explain the most
energetic astrophysical phenomena such as gamma-ray bursts and cosmic
rays from compact objects \citep*{okm,orv}. A phase transition to
quark matter can result in the formation of a quark star, if certain
stability criteria are fulfilled \citep{weber}.

\begin{figure}[t!]
\includegraphics[width=0.5\textwidth]{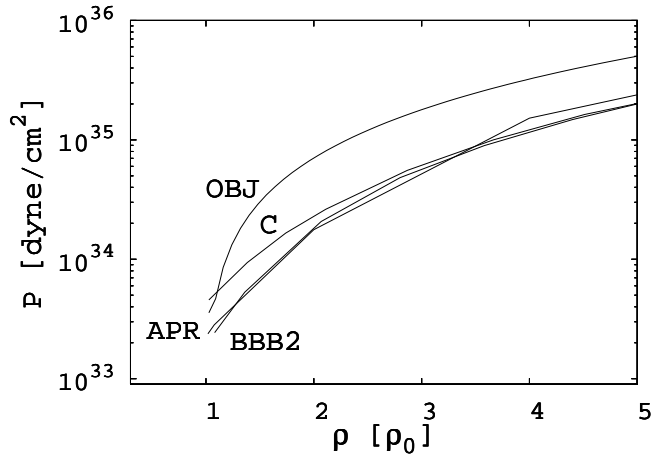}
\caption{The EOS used (see \S 3). \label{eosplot}\vspace{-0.2cm}}
\end{figure}

Our main objective in this letter is to obtain accurate timescales for
spin-down of a neutron star to the deconfinement density. Ultimately,
we would like to know the likelihood of a neutron star to undergo a
phase conversion to a quark star. To this end, we explore 4 different
EOS to determine plausible bounds on masses and spin-periods that are
consistent with a phase transition to quark matter in the star's
spin-down lifetime. Aspects of spin-down by magnetic braking and
subsequent deconfinement are outlined in section~\ref{theorysection},
along with the approximations made in this work. In
section~\ref{eossection}, we elaborate on our computations and the EOS
used (see Fig.~\ref{eosplot}), connecting them to results displayed
in Table \ref{maxmintable5} and
Figure~\ref{nuvsemulti}. Section~\ref{resultsection} gathers our
principal conclusions based upon an interpretation of results in
Table~\ref{spindowntime5} and Figures~\ref{evst5} \&~\ref{Pvst5}.

\section{Spin-down and deconfinement} \label{theorysection}

It is well-known that the energetics of pulsed emission as well as
unpulsed nebular radiation from a neutron star can be generally
understood as deriving from its rotational kinetic energy
\citep{man}. While the rotational power can be calculated from
measurements of the spin-period ($P$) and its derivative ($\dot{P}$)
in a model-independent way, theoretical calculations linking surface
magnetic fields with the spin-down rate (loss of rotational kinetic
energy) require a model for the magnetic field configuration. In this
paper, we assume a dipolar time-independent magnetic field for the
sake of simplicity. Although more complicated configurations with
toroidal components are likely in reality, these will not affect our
quantitative results which depend more on the absolute strength of the
magnetic field rather than its detailed shape. Further, surface
magnetic fields are likely to decay over a million years or so, and we
do not consider this complication here since its qualitative impact
can be easily inferred from our results. We further assume that
gravitational radiation is a negligible contributor to spin down,
i.e. our neutron stars are born axisymmetric and remain as such, so
that the spin-down rate of a neutron star is entirely due to dipolar
magnetic field braking as given by \citep[e.g.][]{deutsch,man}:

\beqy
\label{spindowneq}
\frac{d\Omega}{dt}=
-\frac{2B^2R^6\Omega^3}{3Ic^3}\sin^2\chi
\eeqy
where $B$ is the equatorial surface magnetic field, $I$ is the moment of inertia, $R$ the equatorial radius of the
star, $\Omega=2\pi/P$ its instantaneous angular velocity and $c$ the
speed of light. In equation (\ref{spindowneq}) above, the inclination 
angle of the dipole ($\chi$) has been averaged over a sphere 
representing a population of  
fixed mass and magnetic field ($\langle \sin^2\chi\rangle=2/3$). 
With all quantities expressed in cgs units,
equation(\ref{spindowneq}) can be integrated to yield the time for a
neutron star to spin down to a particular $\Omega$. The additional
(implicit) input that is required to solve for the spin-down time is
the EOS, which determines the $R$ and $I$ for a given $\Omega$.

Computing the spin-down time to deconfinement requires precise
knowledge of the details of a phase transition to quark
matter. Although yet to be determined exactly from first principles,
it is expected to occur on general grounds at high baryonic density,
whose value depends on the EOS employed \citep{HP,GLD}. If the
interface tension between the quark and nuclear phases is too high, a
first order phase transition to a pure quark matter core at very high
density ($\sim 10\rho_0$) is likely, by which time the star may become
gravitationally unstable and collapse to a black hole. Several other
phases, such as meson-condensed phases or structured nuclear matter
can intervene at lower densities, softening the EOS. The concomitant
rapid contraction of the softer matter can lead to non-monotonic
behavior of the angular velocity $\Omega$ and cusps in the braking
index $n=\Omega\ddot{\Omega}/\dot{\Omega}^2$ \citep{heisel}.  It is
also possible that the transition from nuclear to quark matter at high
density is second-order if the interface tension between these phases
is small enough. The transition may then occur smoothly over a range
of densities from $(4-8)\rho_0$ with an increasing proportion of quark
matter in the mixed phase \citep{Glen}, and a continuous variation of
the pressure and angular velocities. Our working assumptions will be
in line with the latter, and we choose a threshold deconfinement
density of $5\rho_0$. Obviously, our quantitative results depend on
this assumed value, but will remain qualitatively true irrespective of
the transition density one may choose. When the central baryonic
density reaches the critical density of $\rho_{\rm crit}=5\rho_0$,
deconfined quark matter will appear in the inner regions of the
neutron star. If the strange quark matter hypothesis \citep{itoh}
holds true, it is possible that the entire star converts to a quark
star in a violent phenomenon termed the Quark Nova \citep{ODD,
KOJ}. Aspects and consequences of the Quark Nova are discussed in
\citet{okm} and \citet{orv}.

\section{Neutron star models and Equations of State} \label{eossection}

\begin{figure}[b!]
\includegraphics[width=0.5\textwidth]{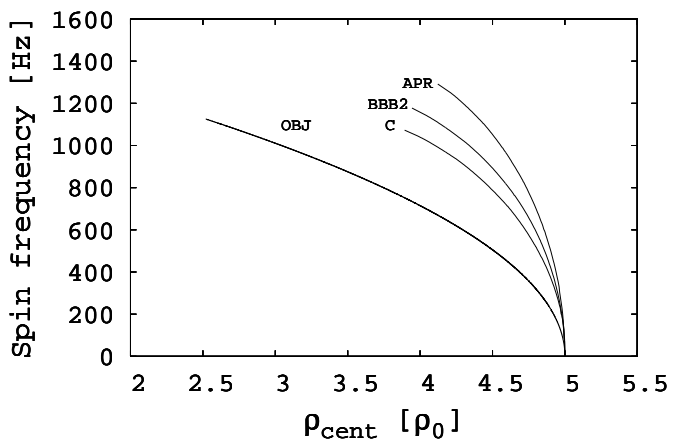}
\caption{Frequency vs central density for EOS used. All sequences are with constant baryonic mass and for minimum mass configurations (see Table~\ref{maxmintable5}). The minimum mass configurations are those  that
 can reach a critical density of $5\rho_0$ due to spin
 down.\label{nuvsemulti}}
\end{figure}

In order to explore the increase in central density due to spin-down,
we create sequences of neutron star models using the RNS code
\citep{SF} developed specifically to treat rapidly rotating neutron
stars. Models in a particular sequence have the same EOS, constant
baryon number, increasing central density and decreasing angular
velocity. We assume that the neutron star is born axisymmetric and
initially rotates at Keplerian frequencies. For each sequence, the
$RNS$ code outputs (among other physical quantities) $I$, $\Omega$,
and $R$, which feed into equation(\ref{spindowneq}). We emphasize that
since the magnetic field does not appear explicitly in the
calculations performed by the RNS code, it does not affect the
structural evolution of the star or the central density at which the
transition to quark matter occurs. In principle, extremely high
magnetic fields exceeding $10^{15}$G can affect neutron star structure
\citep{BQ95}. In our simplified approach, as per
equation(\ref{spindowneq}), the magnetic field only determines the
time to reach the deconfinement density.

We employ 4 different EOS which are, in increasing order of stiffness:
BBB2 \citep{baldo}, C \citep{eosC}, APR \citep*{APR98} and OBJ
\citep{OBJ}. The pressure versus density curve for each of the above
EOS is plotted in Figure~\ref{eosplot} for densities ranging from
$\rho_0$ to $5\rho_0$. The OBJ EOS is a lot stiffer than the rest at
high densities, due to strong repulsion from vector mesons at short
distances \citep[details in][]{OBJ}.

The minimum baryonic mass that is required for a given EOS to support
spin-down to deconfinement is found by the RNS code as the sequence
that spins down to $\rho_{\rm crit}$ at zero spin. The gravitational
mass of such a model at zero spin is what we call minimum mass.
Figure~\ref{nuvsemulti} displays the spin frequency of neutron stars
with minimum mass, as a function of its central density. The curves
shown are for a $1.46M_\odot$ star with EOS BBB2, a $1.53M_\odot$ star
with EOS C, a $1.78M_\odot$ star with EOS APR, and a $2.8M_\odot$ star
with EOS OBJ (all masses are gravitational masses at zero spin).

Equation (\ref{spindowneq}) implies
 that it takes infinitely long to spin-down to zero frequency. In
 practice, we therefore compute the time to reach within $1\%$ of 
 $\rho_{\rm crit}$, as illustrated in Figure~\ref{evst5}. 
For a minimum mass
star, we take this time to be the maximum time for a star with a given
EOS to reach deconfinement density. Neutron stars with mass lower than
the minimum mass will not reach deconfinement density due to
spin-down, and stars with higher masses will reach the critical
density sooner, before they have spun down completely. Conversely, a
maximum mass can also be found, such that any mass beyond this would
already have central densities exceeding $\rho_{\rm crit}$ at the
Kepler frequency so that it is already in a deconfined state and
spin-down causes no further change.

The minimum and maximum masses as described above are listed in
Table~\ref{maxmintable5} for each EOS. Note that the OBJ EOS, on
account of its stiffness, predicts a large minimum mass in order to
spin-down to deconfinement. The minimum density at Keplerian frequency
required to reach deconfinement density ($\rho_{\rm min,K}$) is lowest
for the OBJ EOS and highest for the APR EOS (these are the two
stiffest EOS). This is because the moment of inertia is so much larger
for the OBJ EOS as compared to the others, that it more than
compensates for the increase in angular momentum with increasing
stiffness of the EOS. This also explains the lower spin frequency at a
given central density for the OBJ EOS.

\section{Results and Conclusions}\label{resultsection}

\begin{table}[b!]
\begin{center}
\caption{Columnwise: minimum gravitational mass at zero
spin, maximum gravitational mass at Kepler rotation and minimum
density at Kepler rotation that can reach the critical density
($\rho_{\rm crit}=5\rho_0$) due to spindown.\label{maxmintable5}}
\begin{tabular}{llll}
\hline \noalign{\smallskip} EOS & $M_{\rm min}$ ($M_\odot$) & $M_{\rm
max}$ ($M_\odot$)& $\rho_{\rm min, K} (\rho_0)$ \\
\noalign{\smallskip} \hline \hline \noalign{\smallskip} BBB2 & $1.46$
& $1.79$ & $3.92$ \\ APR & $1.78$ & $2.16$ & $4.12$ \\ C & $1.53$ &
$1.83$ & $3.88$ \\ OBJ & $2.8$ & $3.36$ & $2.52$ \\
\noalign{\smallskip} \hline
\end{tabular}
\end{center}
\end{table}

\begin{figure*}[t!]
\center
\includegraphics[width=0.32\textwidth]{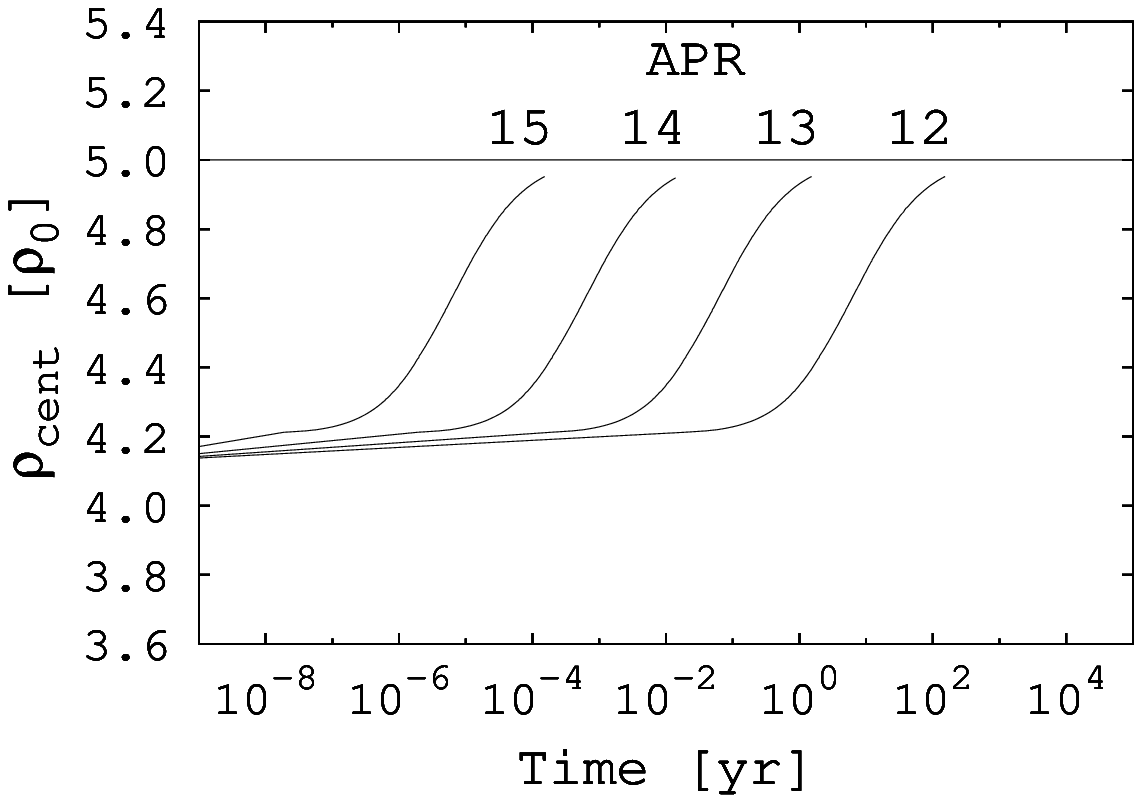}
\includegraphics[width=0.32\textwidth]{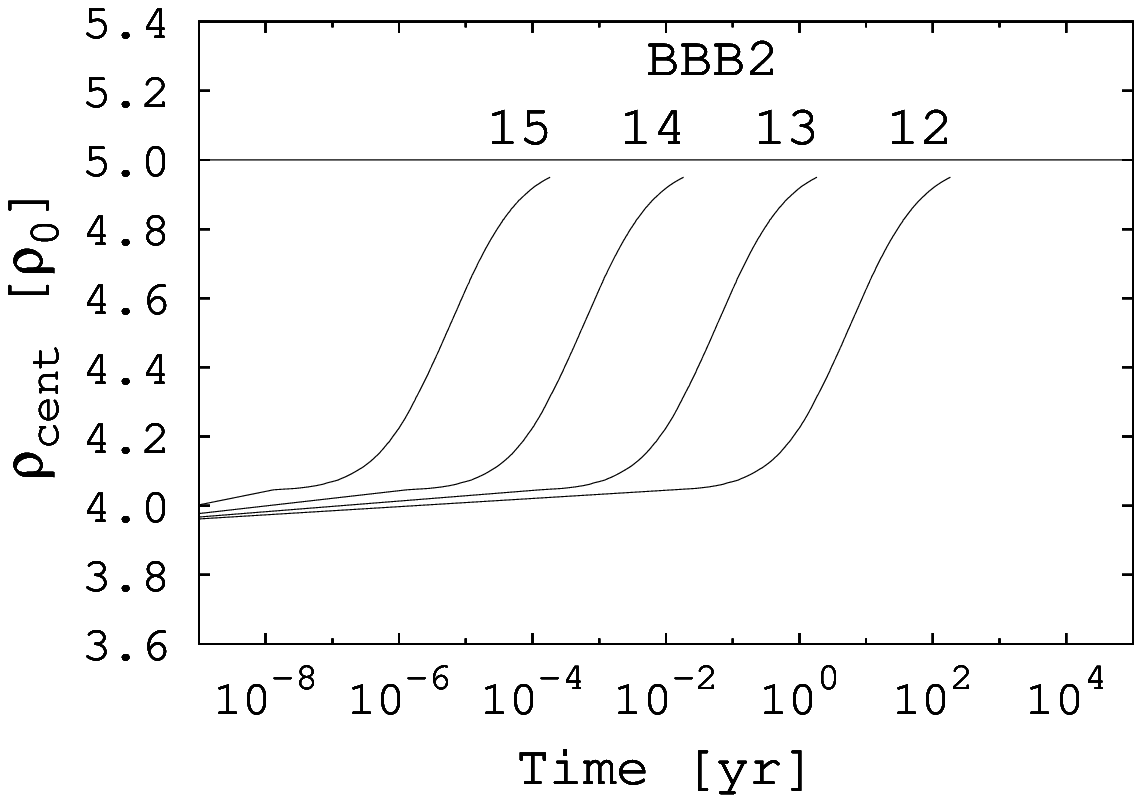}
\\
\includegraphics[width=0.32\textwidth]{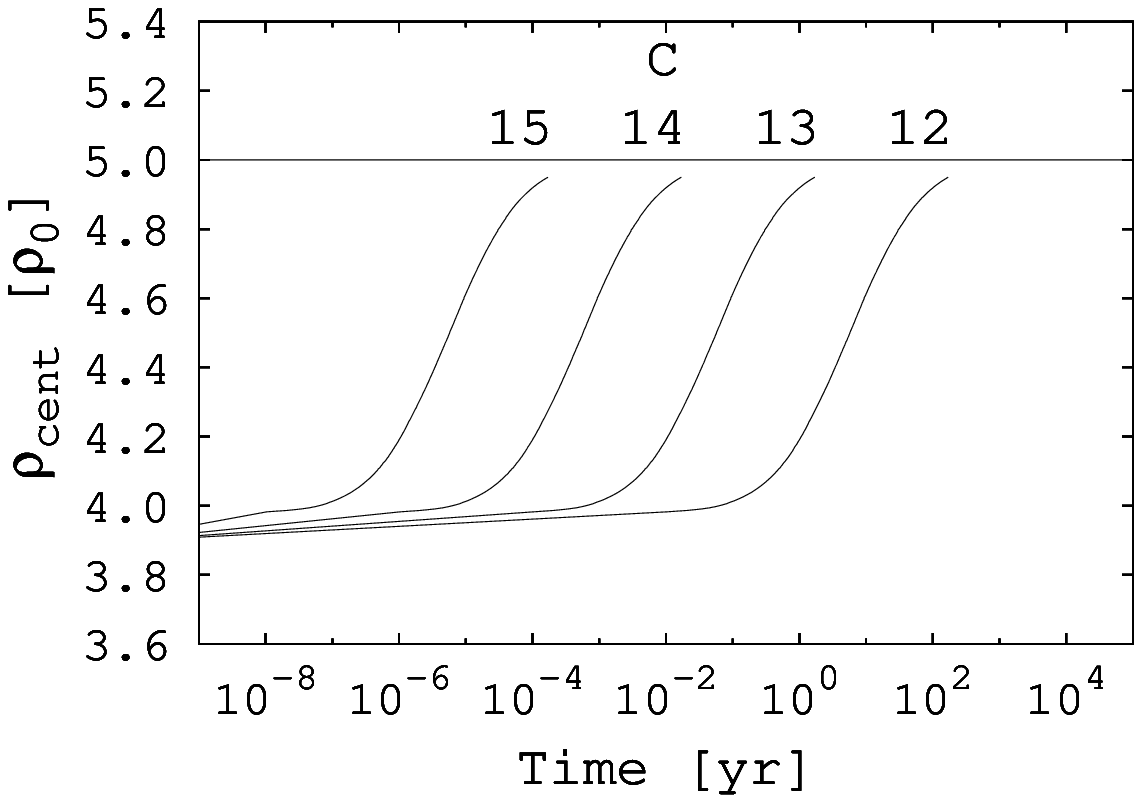}
\includegraphics[width=0.32\textwidth]{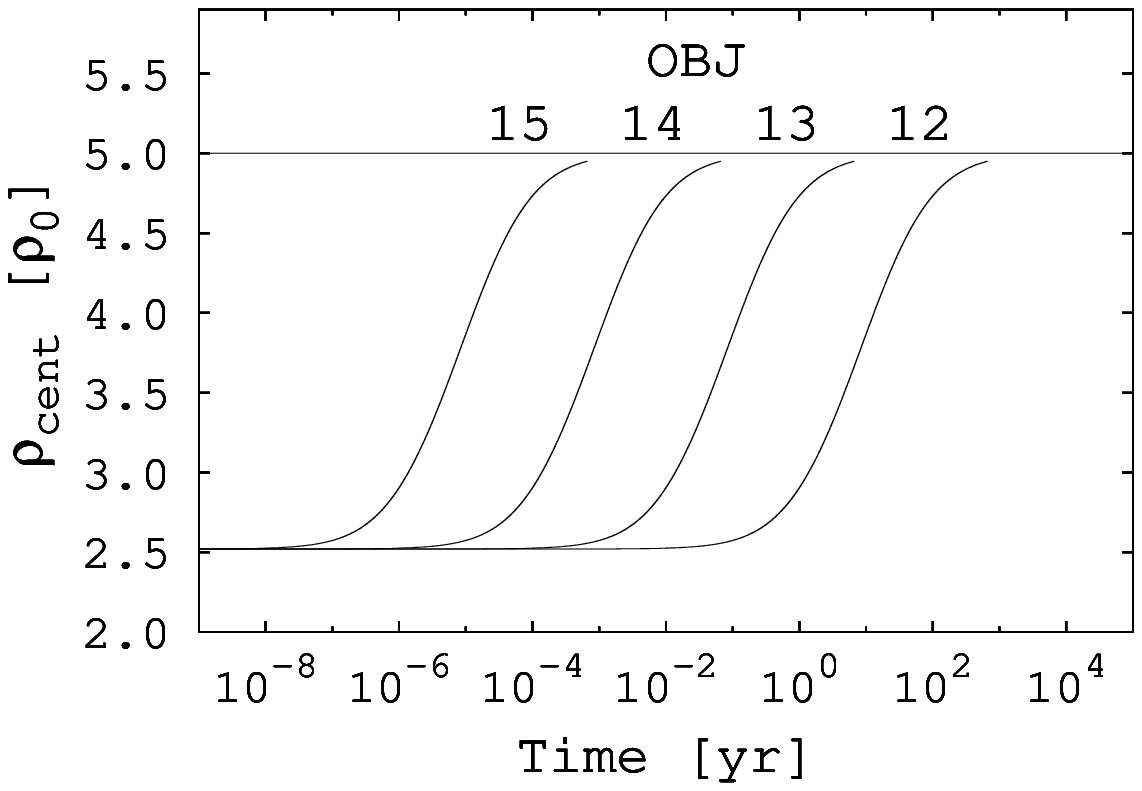}
\caption{Central density versus time for each of
the EOS used with $\rho_{\rm crit}=5\rho_0$. All of these models has
the minimum gravitational mass (see Table~\ref{maxmintable5}). The
magnetic field is $10^X$~G with $X=12, 13, 14, 15$. The curves
terminate when $\rho$ is within $1\%$ of $\rho_{\rm crit}$.}
\label{evst5}
\includegraphics[width=0.32\textwidth]{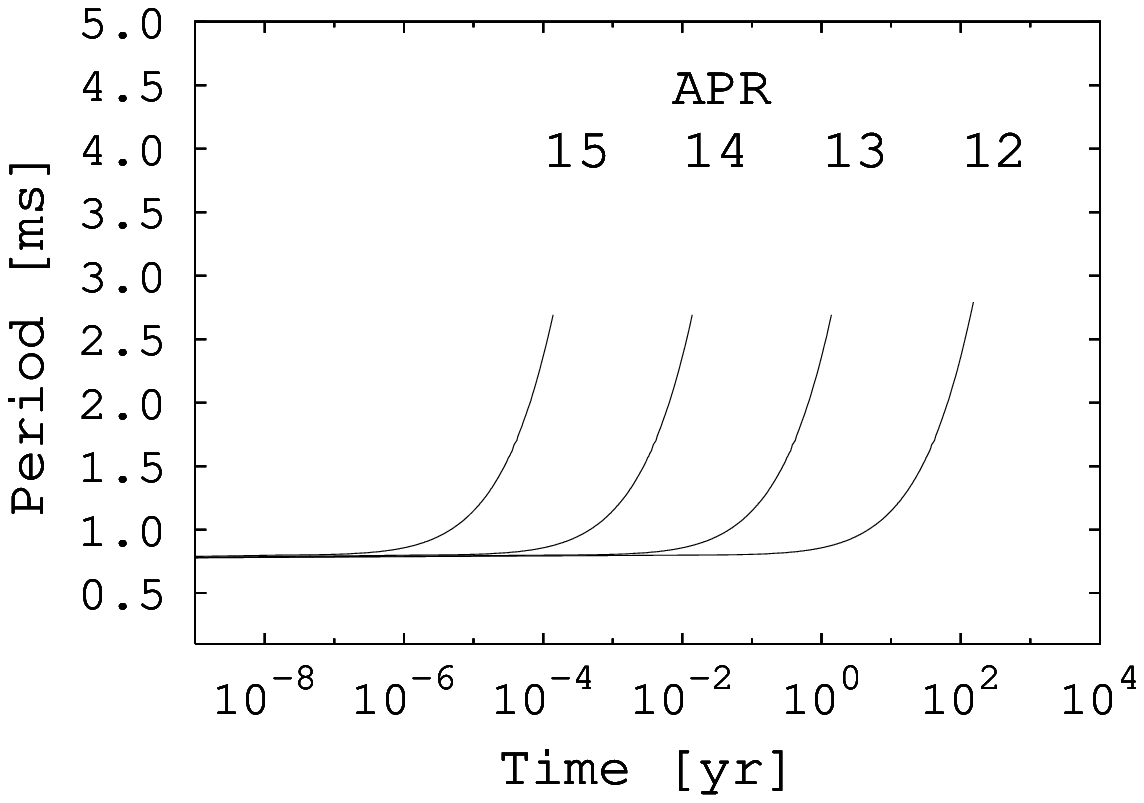}
\includegraphics[width=0.32\textwidth]{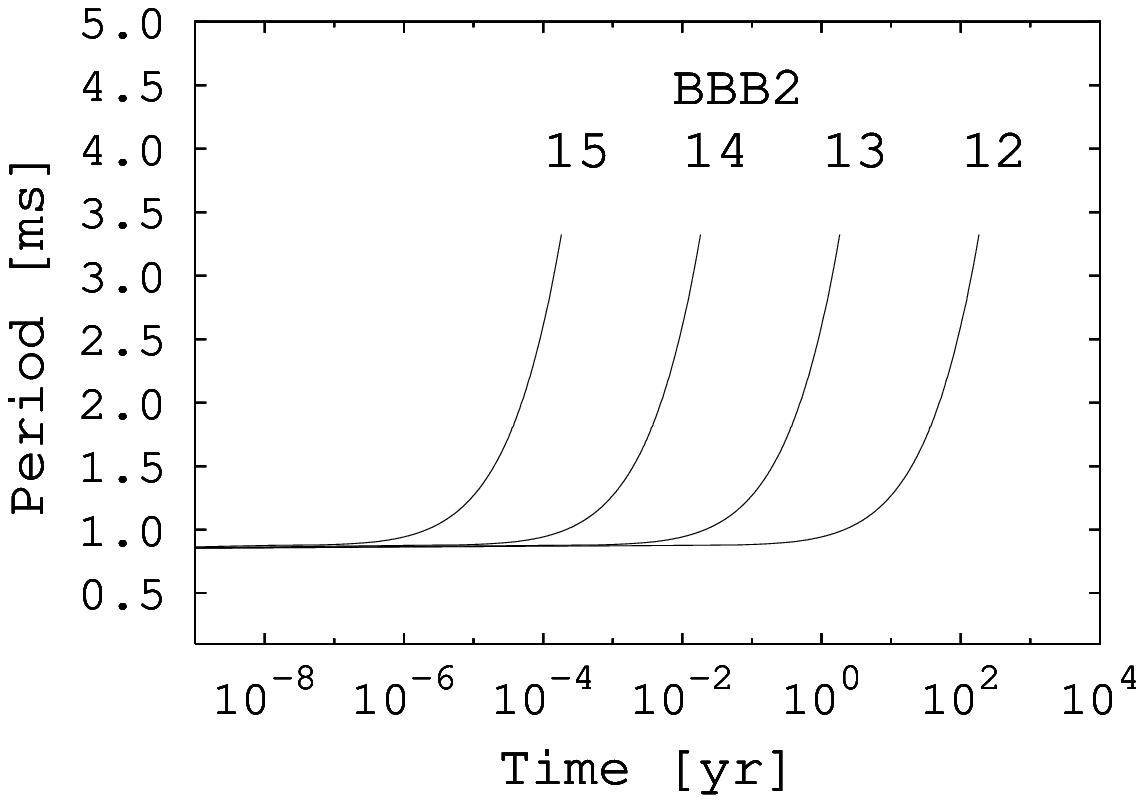}
\\
\includegraphics[width=0.32\textwidth]{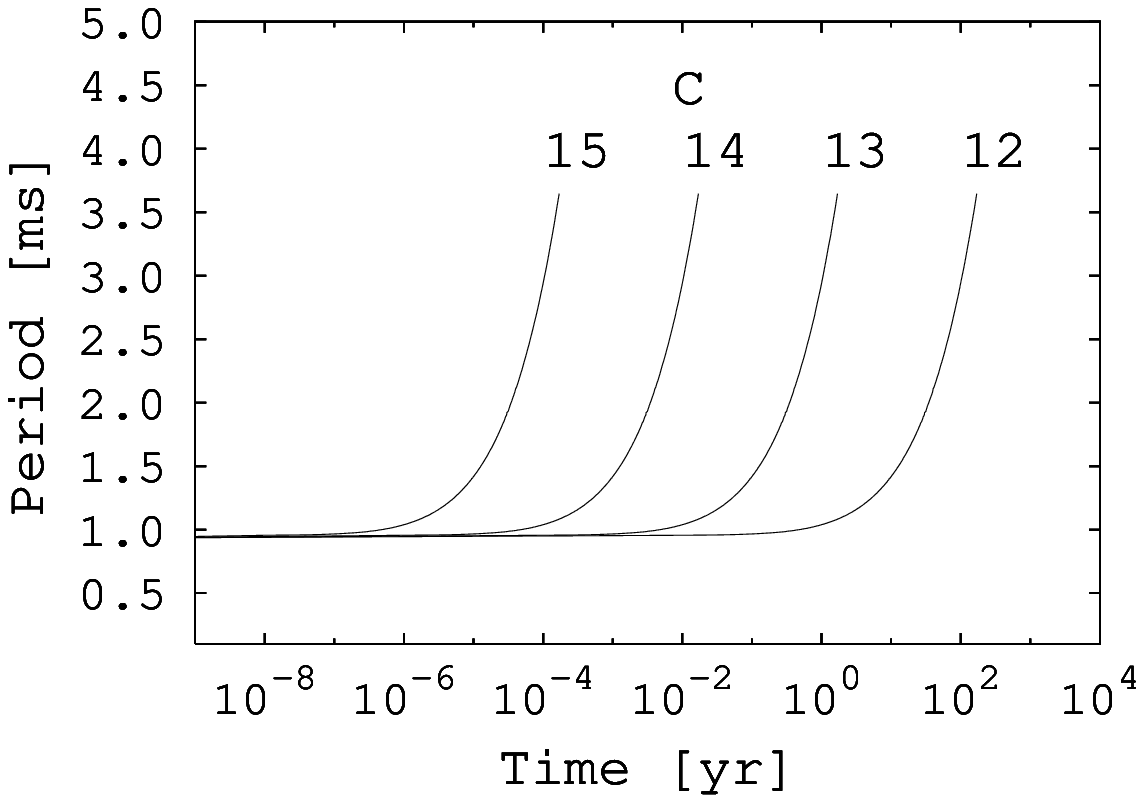}
\includegraphics[width=0.32\textwidth]{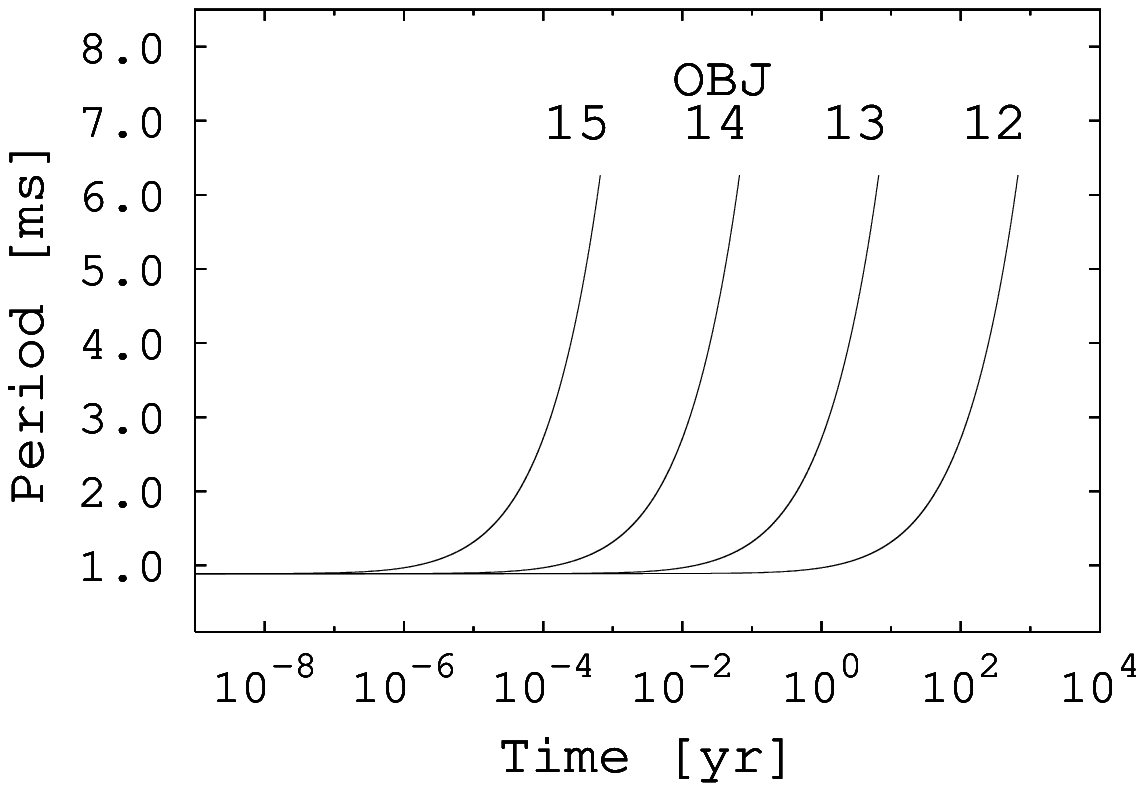}
\caption{Period vs time for the EOS studied in this paper with
$\rho_{\rm crit}=5\rho_0$, with the minimum gravitational masses
listed in Table~\ref{maxmintable5} and for different magnetic
fields. The curves terminate when $\rho$ is within $1\%$ of $\rho_{\rm
crit}$. Note that for a given EOS all curves reach the same final 
period since the magnetic field determines only the 
time to reach deconfinement, and does not enter the hydrostatic equations
(see text).}
\label{Pvst5}
\end{figure*}

The maximum time to reach deconfinement via spin-down can be read off
from Figure~\ref{evst5} for each EOS for magnetic fields in the range
$10^{12}-10^{15}$G. For any of the EOS, it takes up to a few hundred
years
for the star to spin down to $0.99\rho_{\rm crit}$ for a $10^{12}$G
magnetic field. This time is drastically reduced to about an hour for
a $10^{15}$G magnetic field. It is important to emphasize that the
curves in Figure~\ref{evst5} correspond to the minimum mass star in
each case, i.e. it takes an infinite time for the magnetic field to
spin such a star down to $\rho_{\rm crit}$. In
Table~\ref{spindowntime5}, we therefore list the time it takes the
star to increase its central density to within $1\%$ of the critical
density (i.e $0.99\rho_{\rm crit}$) due to spin down.

\begin{table}[b!]
\begin{center}
\caption{Time to spin down from Kepler frequency to the frequency 
corresponding to $1\%$ of $\rho_{\rm
crit}$ for a star with minimum mass for different magnetic field 
strength.\label{spindowntime5}}
\vskip -0.3cm
\begin{tabular}{llllll}
\hline \noalign{\smallskip} & & & Time & \\ EOS & $M_{\rm min}$
($M_\odot$) & $10^{15}$ G & $10^{14}$ G & $10^{13}$ G & $10^{12}$ G \\
\noalign{\smallskip} \hline \hline \noalign{\smallskip} BBB2 & $1.46$
& $96$ min & $6.7$ days & $667$ days & $183$ years\\ APR & $1.78$ &
$73$ min & $5.1$ days & $505$ days & $152$ years\\ C & $1.53$ & $89$
min & $6.2$ days & $619$ days & $170$ years\\ OBJ & $2.8 $ & $348$ min
& $24$ days & $2418$ days & $663$ years\\ \noalign{\smallskip} \hline
\end{tabular}
\end{center}
\end{table}

Figure~\ref{Pvst5} shows the period as a function of time for each EOS
and for different magnetic field strengths. Just as before, this is
for minimum mass stars and we plot the period until the central
density is within $1\%$ of $\rho_{\rm crit}$. It is noteworthy that
the central density does not change much after the rotation period
slows to about $P_{\rm max} = 3$ms (about 6 ms for the OBJ EOS). This
is independent of the magnetic field since that only determines the
time to deconfinement, not the evolution of period with time. From the
above limiting value of the period, and the minimum mass results
tabulated in Table~\ref{maxmintable5} for our fiducial critical
density of $5\rho_0$, we conclude that only stars with mass greater
than about 1.5M$_{\odot}$ that have a spin period $P<P _{\rm max}$ can
reach a deconfined phase. Conversely, if the star has $P>P_{\rm max}$
as observed now, it cannot change its central density sufficiently to
undergo deconfinement due to spin down, and must either remain a
neutron star or have been born as a quark star.

Although the critical density and consequently $P_{\rm max}$ is not
known precisely, the range of currently observed isolated neutron star
masses in conjunction with our study implies that such stars can
spin-down to a deconfined phase only if the deconfinement density is
not too much above $5\rho_0$. If the deconfinement density is much
larger than this, neutron stars in the currently observed mass range
\citep{stairs}
cannot increase their central densities enough via spin-down to reach
the threshold, and will remain neutron stars for their lifetimes. This
conclusion can only be avoided by assuming that all neutron stars are
actually quark stars and are born as such in the aftermath of
supernova explosions. Since most realistic quark matter EOS cannot
support a maximum mass greater than 1.6M$_{\odot}$ \citep[see
e.g.][]{dey98}, it is unlikely that this is true. We therefore believe
that the existence of quark stars is predicated upon having a low
deconfinement threshold ($\leq 5\rho_0$). Once this threshold is
pinned down via theoretical studies addressing Quantum Chromodynamics
at high baryon
density, we will be able to make more definitive statements about the
existence of quark stars and related phenomena. Until then, the
possibility that some isolated neutron stars are really quark stars
cannot be dismissed.

\acknowledgments{
We thank S. Morsink for help with the RNS
code and Ken Nollett for informative discussions.
The research of R.O. is supported by an operating grant from the
Natural Science and Engineering Research Council of Canada (NSERC) as
well as the Alberta Ingenuity Fund (AIF). J.S. acknowledges the 
hospitality of the University of Alberta and of Argonne National
Laboratory where parts of this work were performed. P.J. is supported
by the Department of Energy, Office of Nuclear Physics, contract
no. W-31-109-ENG-38.}

\end{document}